\documentclass[iop]{emulateapj}
\usepackage{amsmath}

\usepackage{color}

\shorttitle{}
\shortauthors{}

\begin{document}


\title{Chaos in the Test Particle Eccentric Kozai-Lidov Mechanism}
\author{Gongjie Li \altaffilmark{1}, Smadar Naoz \altaffilmark{1}, Matt Holman \altaffilmark{1}, Abraham Loeb \altaffilmark{1}}
\affil{$^1$ Harvard-Smithsonian Center for Astrophysics, The Institute for Theory and
Computation, \\60 Garden Street, Cambridge, MA 02138, USA}
\email{gli@cfa.harvard.edu}



\begin{abstract}
The Kozai-Lidov mechanism can be applied to a vast variety of astrophysical systems involving hierarchical three-body systems. Here, we study the Kozai-Lidov mechanism systematically in the test particle limit at the octupole level of approximation. We investigate the chaotic and quasiperiodic orbital evolution by studying surfaces of section and the Lyapunov exponents. We find that the resonances introduced by the octupole level of approximation cause orbits to flip from prograde to retrograde and back as well as cause significant eccentricity excitation, and the chaotic behaviors occur when the mutual inclination between the inner and the outer binary is high. We characterize the parameter space that allows large amplitude oscillations in eccentricity and inclination.
\bigskip
\end{abstract}


\section{Introduction}
The Kozai-Lidov mechanism \citep{Kozai62, Lidov62} has proven very useful for interpreting numerous astrophysical systems. For example, it has been shown that it can play a major role in exoplanet configurations and obliquities \citep[e.g.][]{Holman97, Wu03, Fabrycky07, Veras10, Correia11, Naoz11, Naoz12}. In addition, close stellar binaries with two compact objects are likely produced through triple evolution, and the Kozai-Lidov mechanism may play a key role in these systems (e.g. \citealt{Harrington69, Mazeh79, Soderhjelm82, Kiseleva98, Ford00, Eggleton01, Fabrycky07, Perets09F, Thompson11, Katz12, Shappee13, Naoz13}; Naoz \& Fabrycky submitted). Furthermore, the Kozai-Lidov mechanism has been proposed as an important element in the growth of black holes at the centers of dense star clusters, the formation of short-period binaries black hole \citep{Blaes02, Miller02, Wen03, Ivanova10}, and tidal disruption events (\citet{Chen09, Chen11, Wegg11, Bode13}, Li et al., in prep).

The Kozai-Lidov mechanism focuses on hierarchical three-body systems, which can be treated as the interaction between two elliptical wires by orbit averaging: the inner wire is composed of the inner two objects, and the outer wire is composed of the outer companion orbiting around the center mass of the inner two objects. The total angular momentum of this system, the vector sum of the inner orbit's and the outer orbit's angular momenta, is conserved. 

\citet{Kozai62} and \citet{Lidov62} first studied this mechanism by expanding the gravitational potential in a power series of the semi-major axis ratio and considered applications when one of the inner object is massless  (the test particle limit) and the outer orbit is circular. \citet{Kozai62} considered the secular (long term) evolution of asteroids under the perturbation of Jupiter, and \citet{Lidov62} studied the secular evolution of satellites under the perturbation of the Moon. In those cases, the gravitation potential of the inner orbit is axisymmetric, which renders the $\hat{z}$ component of the inner orbit's angular momentum ($J_z$) constant, where $\hat{z}$ is the direction of the total angular momentum of the system. The quadrupole order of approximation ($O\big((a_1/a_2)^2\big)$) sufficiently describes the orbital evolution of such systems, and the eccentricity and the inclination undergo large amplitude oscillations due to the ``Kozai resonance" when $i>39.2^\circ$. 

Recently, \citet{Naoz11} considered the case when none of the inner objects is a test-particle, and pointed out that $J_z$ is no longer conserved. In addition, the eccentric Kozai-Lidov Mechanism (hereafter EKL) applies to cases when the outer orbit is non-circular, where the $\hat{z}$ component of the angular momentum of the inner orbit is also not conserved \citep{Naoz11}. In this situation, the octupole terms in the potential ($O\big((a_1/a_2)^3\big)$) need to be taken into account to describe the orbital evolution, where the eccentricity of the inner orbit can be excited to unity, and the inner orbit may flip from prograde to retrograde or vice versa \citep{Naoz11, Lithwick11, Katz11, Naoz13}. As the eccentricity increases, the pericenter distance decreases, and causes an enhanced tidal disruption rate (Li et al., in prep). Furthermore, including the octupole effects, the oscillation in the eccentricity and the inclination of the inner orbit may still exist when $i<39.2^\circ$, and the inner orbit may undergo a coplanar flip from $\sim0^{\circ}$ to $\sim180^{\circ}$ \citep{Li13}.

Here, we probe the test particle limit, which simplifies the analysis due to its smaller number of degrees of freedom. This approximation was proven to be very useful in a large range of astrophysical settings (\citet{Lithwick11, Katz11, Naoz12, Li13}, Naoz \& Silk in prep, Li, et al., in prep). Importantly, probing this limit can help us gain some basic understanding of the EKL mechanism. The test particle limit has been studied in the literature before to obtain an analytical understanding on the flip of the orbit \citep{Lithwick11, Katz11}. Nevertheless, a systematic study on the chaotic behavior and the identification of the underlying resonances are necessary but are uncovered in the literature. We identify the resonances, and characterize the chaotic regions and the initial conditions where high eccentricity and the flips may occur in the parameter space. This can help predict the dynamical evolution of systems without doing a large amount of simulations. 

This paper is organized as follows. In \textsection 2, we give a brief overview of the Kozai-Lidov mechanism. In \textsection 3, we investigate the surface of section systematically for a large range of orbital parameters. In \textsection 4, we characterize the initial condition which allows large amplitude oscillations in eccentricity and inclination. Finally in \textsection 5, we characterize the chaotic regions.

\section{Overview of the eccentric Kozai-Lidov mechanism in the test-particle limit}
\begin{figure}
\includegraphics[width=4in, height=3in]{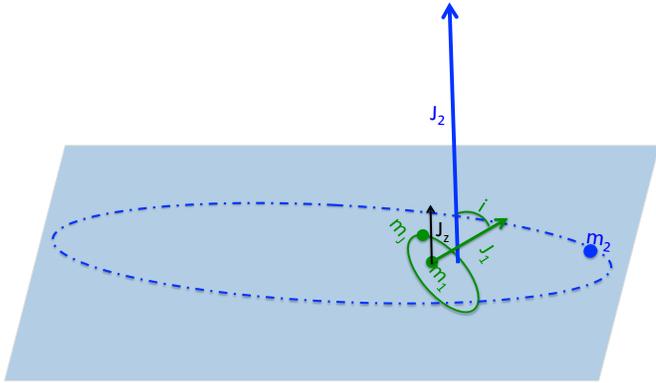}
\caption{\label{f:config} The system configuration. A test particle $m_t$ orbits around an object $m_1$ and forms the inner binary. The outer binary consists of the object ($m_2$) and $m_1$ (in the test particle limit). $J_o$ represents the angular momentum of the outer binary, $J$ represents that of the inner binary, and $J_z$ represents the $\hat{z}$ component of $J$, where $\hat{z}$ is in the direction of $J_o$. In the test particle limit $J \ll J_o$ and the outer orbit is stationary.}
\vspace{0.1cm}
\end{figure}

As mentioned in the introduction, the Kozai-Lidov mechanism describes the dynamical behavior of hierarchical three-body systems (see Figure \ref{f:config}). The inner two objects ($m_1$ and $m_t$) form an inner orbit, and the outer orbit is formed by the outer object ($m_2$) orbiting around the center mass of the inner two objects. The eccentric Kozai-Lidov mechanism describes the dynamics when the outer orbit is eccentric, and the test-particle limit requires one of the closely separated objects to be a test particle $m_t \to 0$. 
  
In the hierarchical configuration, we average over the mean motion of the two orbits and treat the evolution of the system as the interaction of two elliptical wires known as the secular approximation. This approximation reduces this system from six degrees of freedom to four degrees of freedom. In addition, in the test-particle limit, the outer orbit is stationary, and reduces the system to two degrees of freedom \citep{Harrington68, Harrington69, Ford00}. Expanding the Hamiltonian of the interaction energy between the two ellipses in a power series of $a_1/a_2$, the Hamiltonian can be expressed as the following at the second (quadrupole) and the third (octupole) order  \citep{Lithwick11}: 

\begin{align}
F_{quad} (J, \omega, J_z, \Omega) &= \frac{1}{2}(-1+ J^2) + \frac{J_z^2}{J^2}+\frac{3(1 - J^2) J_z^2}{2 J^2} \\ \nonumber
 &+\frac{1 - J^2}{1 - J_z^2/J^2}\cos{(2 \omega)} \\
F_{oct} (J, \omega, J_z, \Omega) &= \frac{5}{16}(\sqrt{1-J^2}+\frac{3}{4}(1-J^2)^{3/2})\\ \nonumber
&\Big[(1-\frac{11J_z}{J} -\frac{5J_z^2}{J^2}+\frac{15 J_z^3}{J^3}) \cos{(\omega - \Omega)} \\ \nonumber
&+ (1+ \frac{11 J_z}{J} -\frac{5 J_z^2}{J^2} -\frac{15 J_z^3}{J^3}) \cos{(\omega + \Omega)}\Big] \\ \nonumber 
&- \frac{175}{64}(1-J^2)^{3/2} \\ \nonumber 
&\Big[(1-\frac{J_z}{J}-\frac{J_z^2}{J^2}+\frac{J_z^3}{J^3})\cos{(3\omega-\Omega)} \\ \nonumber 
&+ (1+\frac{J_z}{J}-\frac{J_z^2}{J^2}-\frac{J_z^3}{J^3})\cos{(3 \omega + \Omega)}\Big] ,
\end{align}
where $H_{quad} = -F_{quad}$ and $H_{oct} = -F_{quad} - \epsilon F_{oct}$, and
\begin{equation}
\epsilon = \frac{a_1}{a_2}\frac{e_2}{1-e_2^2} .
\end{equation}
$\epsilon$ characterizes the importance of the octupole order. The Hamiltonian is scaled with $m_t\sqrt{Gm_1a_1} t_K$, where 
\begin{equation}
t_K=\frac{8}{3}P_{in}\frac{m_1}{m_2}\Big(\frac{a_2}{a_1}\Big)^3(1-e_2^2)^{3/2}
\label{eqn:tk}
\end{equation}
\citep{Lithwick11}. $J=\sqrt{1-e_1^2}$ is the specific angular momentum of the inner orbit, $\omega$ is the argument of periapsis of the inner orbit, $J_z = \sqrt{1-e_1^2}\cos{i_1}$ is the $\hat{z}$ component of the inner orbit's angular momentum $J$, and $\Omega$ is the longitude of the ascending node of the inner orbit. Specifically, $J$, $\omega$ and $J_z$, $\Omega$ are conjugate momentum and coordinate pairs. We denote $e_1$ as the eccentricity of the inner orbit, and $i_1$ as the inclination of the inner orbit to the total angular momentum of the system. In the test particle limit, $i_1 = i$ is the mutual inclination between the two orbits. 

The secular approximation breaks down when the change in the angular momentum of the inner binary happens faster than the orbital timescales (see eqn B14 in \citet{Ivanov05}, eqn 18 in \citet{Antonini14}, and eqn 48 in \citet{Bode14}). Since the analysis presented here is for the reduced Hamiltonian (note that $J$ is the specific angular momentum: $J=\sqrt{1-e^2}$), which is independent of the masses, the timescales can be set arbitrarily and the secular approximation is irrelevant here. Nevertheless, applying this study to physical systems requires the correct scaling of the masses to satisfy the criteria for the secular approximation.

In the quadrupole limit, the Hamiltonian is independent of $\Omega$, so $J_z$ is constant, and the system is integrable. In addition, the angle $\omega = \varpi - \Omega$ is the resonant angle of the system, where $\varpi$ is the longitude of the periapsis. When $i>39.2^\circ$, the solution admits a resonant region and $e_1$ and $i$ exhibit large amplitude oscillations. Particularly, $e_1$ may be excited to high values starting from $e_1 \sim 0$ \citep[e.g.][]{Morbi02}.

As mentioned in the introduction, the octupole order adds variations in $J_z$ which allows the inner orbit to flip from prograde to retrograde, and the eccentricity to be excited very close to 1 \citep{Lithwick11, Katz11, Naoz11, Naoz12, Naoz13}.  We work with the Hamiltonian at the octupole level of approximation to analyze the surface of section and the chaotic behaviors in the next sections.

\section{Surfaces of section}
\label{s:SS}
\begin{figure}[h]
\includegraphics[width=3in, height=2.25in]{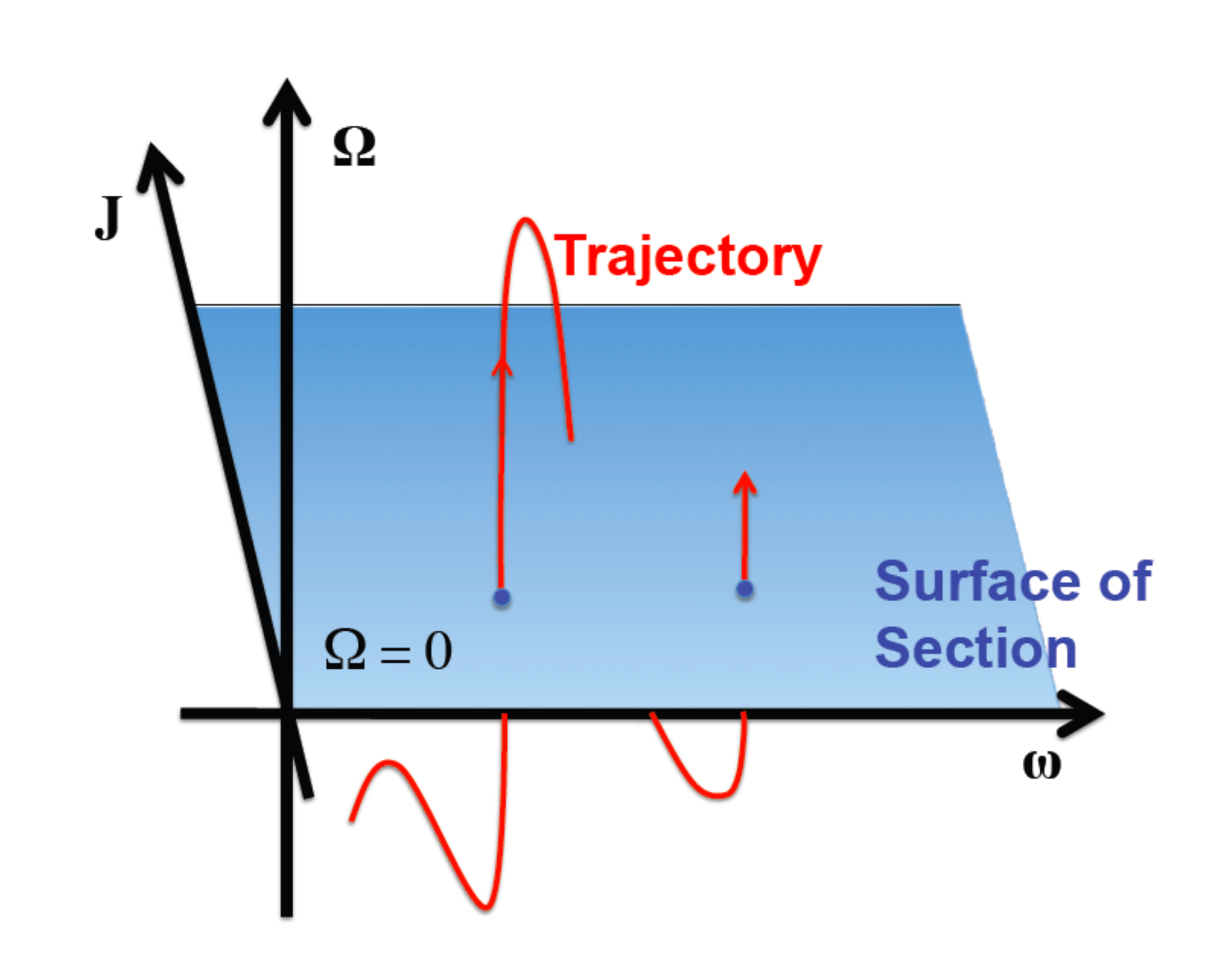}
\includegraphics[width=3in, height=2.25in]{Cartoon2.pdf}
\caption{\label{f:suf} Upper panel: Illustration of the ``surface section" for the $J-\omega$ plane. By recording the point in the trajectory every time $\Omega = 0$, $\dot{\Omega} > 0$, the trajectory can be represented by a 2 dimensional graph, as shown in the left panel. This set of points form the ``surface of section". Lower panel: Illustration of the resonant and chaotic regions in surface of section. We set $H = -0.1$, $\epsilon = 0.1$ in this plot. The resonant and higher order resonant zones are marked by the red and the green arrow. The chaotic zones are indicated by the grey arrow. In the resonant region, the angle $\omega$ is constrained in a small region and the trajectories are quasiperiodic. In the chaotic region, the position of the points are not regular and the trajectories are chaotic.}
\vspace{0.1cm}
\end{figure}

For a two degree of freedom system, the surface of section projects a 4-dimensional trajectory on a 2-dimensional surface. Specifically, we plot points on a 2-dimensional surface composed of one canonically conjugate pair (e.g. $J-\omega$ or $J_z-\Omega$) whenever the other angle ($\Omega$ or $\omega$) reaches a fixed value and moves in a fixed direction (see the left panel in Figure \ref{f:suf}). The collection of the points form the surface of section. 

There are three distinct regions in the surface of section: ``resonant regions", ``circulation regions", and ``chaotic regions" (right panel in Figure \ref{f:suf}). The resonant regions are formed by points where the momenta and coordinates (the angles) undergo bounded oscillations. The trajectories in this region are quasiperiodic, where the system is in the {\it liberation} mode. The {\it circulation} region represents trajectories where the coordinates are not constrained to a specific interval. Both resonant and circulatory trajectories map onto a 1D manifold on the surface of section. On the contrary, chaotic trajectories map onto a 2D manifold. In other words, while quasi-periodic trajectories form lines on the section, chaotic trajectories are area-filling. Embedded in the chaotic region, the small islands correspond to the higher order resonances, which are caused by the interaction between the primary resonances. The trajectories in the higher order resonant regions are also quasiperiodic.

We now consider the surface of section in the $J-\omega$ plane (setting $\Omega = 0$ and $d\Omega / dt > 0$). When $e_1$ is excited to large values, $J \to 0$. When $\Omega$ is set, for each point in the $J-\omega$ plane, $J_z$ ($-J \leqslant J_z \leqslant J$) is unequivocally defined by the conservation of H. There is a finite range of H that the system can take on, because both actions must have zero imaginary components. Since we plot the sections with constant $H$ values, we first explore the range of energy $H$ it can achieve in the $J-\omega$ plane. This way, we can select the range in $H$ that we explore below. 

We notice that the maximum and minimum energy it can reach in the $J-\omega$ plane when $\Omega = 0$ is $\sim 3$ and $\sim-2.4$ (see Appendix Figures \ref{f:maxminHjo} and \ref{f:maxminHjzO}, which show the maximum and the minimum $H$ in the $J-\omega$ plane). Thus, we plot six surfaces of section for $H$ ranging from $H = -2$ to $H=1.2$, since when $H>1.2$, the behavior is similar to that of $H=1.2$. Note that the $H$ admits positive values for this bounded system, because it is the interaction energy between the test particle ($m_t$) and the outer companion ($m_2$), i.e. the disturbing function of this system to the Kepler Hamiltonian of the inner and outer orbits. To investigate the role of the octupole effects, we plot the surface section for two extreme values of $\epsilon$: $\epsilon = 0.001$ and $\epsilon = 0.1$. When $\epsilon<0.001$, the octupole effects are negligible. On the other hand, $\epsilon=0.1$ represents the maximal octupole effects, where when $\epsilon>0.1$, the hierarchical condition may break down and the system may become unstable.

\begin{figure*}[h]
\includegraphics[width=6in, height=2.7in]{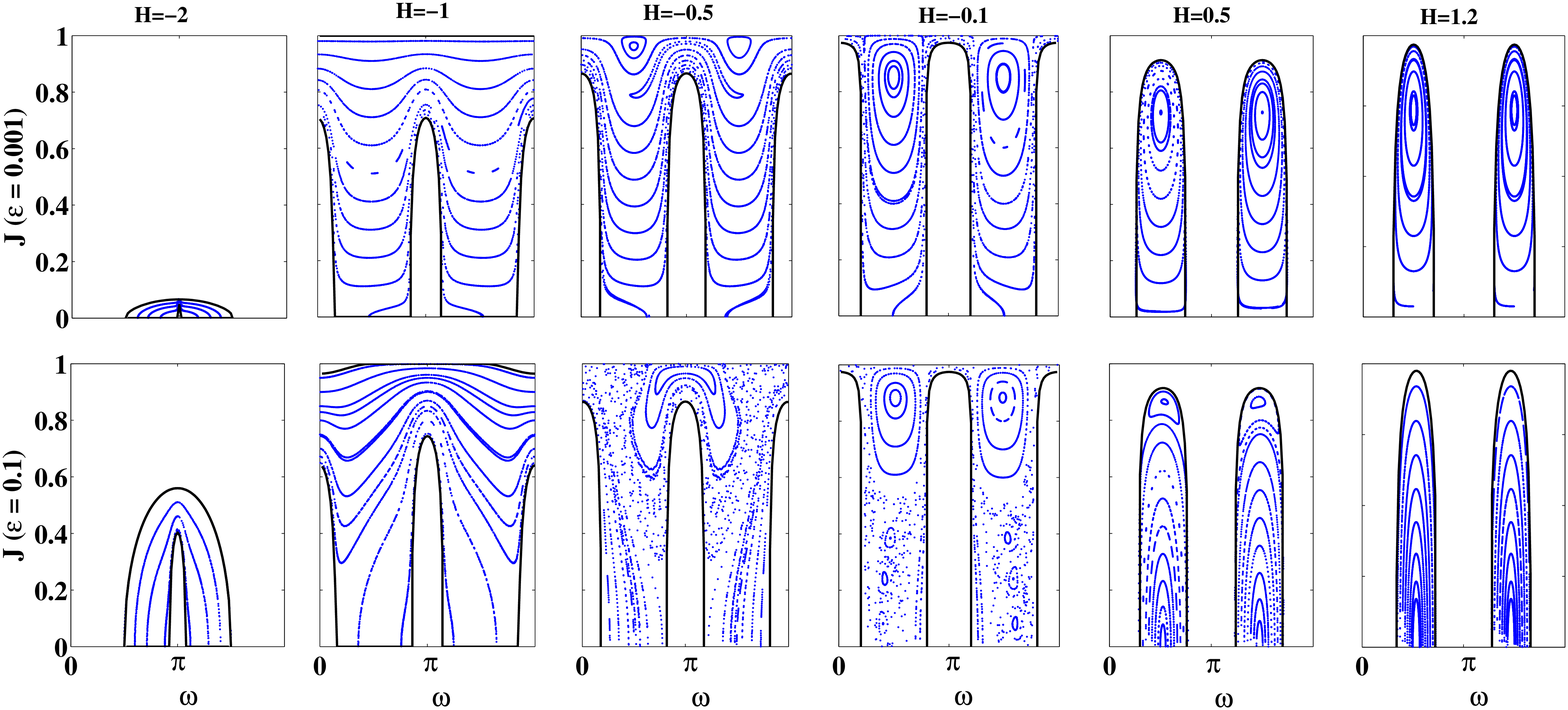}
\caption{\label{f:SSjo} The surface of section in the $J-\omega$ plane. In the first row, $\epsilon = 0.001$ and in the second row, $\epsilon = 0.1$. The octupole terms are important when $\epsilon$ is bigger. $H$ varies from $-2 \sim 1$. The corresponding $e_1$ and $i$ in this plane is shown in Figure \ref{f:joICe} and \ref{f:joICi}. There are chaotic regions at $H = -0.5$ and $H=-0.1$.}
\vspace{0.1cm}
\end{figure*}

The sections are shown in Figure \ref{f:SSjo}. The empty region (bounded by the black curves) do not have physical solutions. The comparison between the two rows in Figure \ref{f:SSjo} shows the difference between the octupole and the quadrupole resonances: $\epsilon = 0.001$ is dominated by the quadrupole effect and $\epsilon = 0.1$ is dominated by both the quadrupole and the octupole effects. For the former, where the quadrupole dominates, there are two resonant regions with fixed points at $\omega = \pi/2$ and $3\pi/2$ when $H$ is high (as shown in Figure \ref{f:SSjo} at $\epsilon = 0.001$, $H = -0.5, -0.1, 0.5$ and $1.2$). For the latter when the octupole plays an important role (i.e., $\epsilon = 0.1$), we find different resonant regions for different energy levels, and the location of the resonant regions vary according to the energy levels. 

The resonant regions are associated with fixed points at $\omega = \pi$, $\omega=\pi/2$ and $\omega=3\pi/2$ depending on the energy level. The resonances at high $J$ and at $\omega = \pi/2$ or $3\pi/2$ correspond to the quadrupole resonances identified in the literature \citep{Kozai62, Holman97, Morbi02}. The other resonant zones result from the interaction of the resonances associated with the ``harmonics" in the octupole level Hamiltonian, i.e. $2\omega$, $\omega \pm \Omega$, and $3\omega \pm \Omega$. Moreover, chaotic regions can only be seen for high $\epsilon$ at $H = -0.5$ and $H = -0.1$, where reading from the surfaces, the chaotic zones are a result of the overlap of the resonances between the quadrupole and the octupole resonances \citep[e.g.][]{Chirikov79, Murray97}. Embedded in the chaotic region, higher order resonances can be found at $H=-0.1$, where the trajectories are quasi-periodic and the eccentricity cannot be excited. 

On the other hand, the comparison between the different energy levels shows the orbital evolution corresponds to different orbital parameters. The corresponding $e_1$ and $i$ are shown in Figure \ref{f:joICe} and Figure \ref{f:joICi} in the appendix. Accordingly, the low $H$ corresponds to the low inclination ($i\sim 0 - 30^\circ$) and high eccentricity ($e_1 \gtrsim 0.6$) case, the higher $H$ corresponds to the high inclination ($i\sim 30^\circ - 60^\circ$) and low eccentricity ($e_1 \lesssim 0.6$) case, and $H>0$ corresponds to high inclination ($i\sim 60 - 90^\circ$) and low eccentricity case ($e_1 \lesssim 0.3$). When $H$ is low ($H \sim -2$), the evolution is only affected by the octupole resonances, while when $H$ is higher, octupole and quadrupole resonances both contribute and may overlap to cause the chaotic region as mentioned above. We find that $e_1$ can be excited to high values ($J \to 0$) for almost all energy levels but is only excited very close to unity for higher $\epsilon$. This emphasizes that the octupole level of approximation causes large eccentricity excitation, since larger $\epsilon$ implies that the octupole level is important. 

\begin{figure*}[h]
\includegraphics[width=6in, height=2.7in]{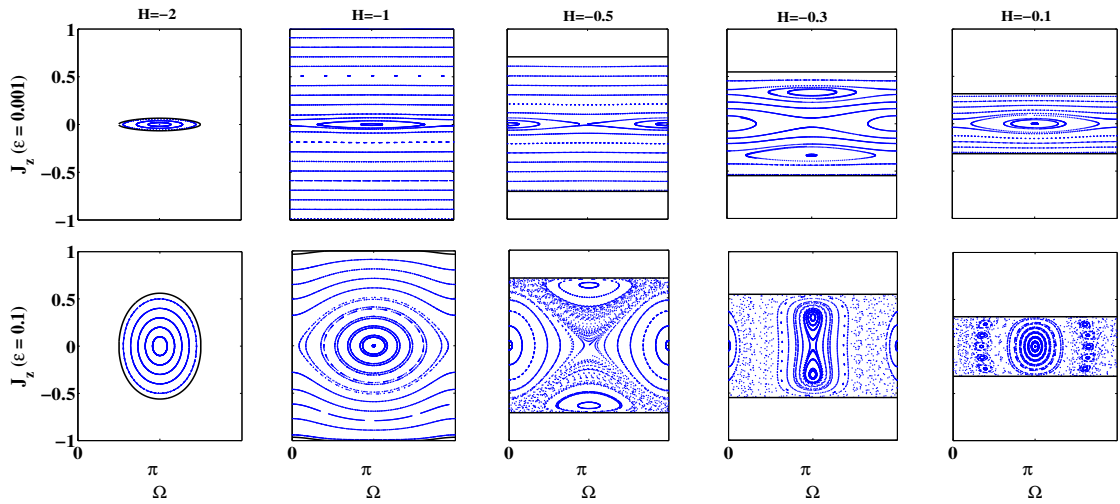}
\caption{\label{f:SSjzO} The surface of section in the $J_z-\Omega$ plane. In the first row, $\epsilon = 0.001$ and in the second row, $\epsilon = 0.1$. $H$ varies from $-2 \sim 0$. There are chaotic regions at $H = -0.5$, $H = -0.3$ and $H=-0.1$. All the features are due to the octupole order, as the $J_z$ is constant in the quadrupole order. The corresponding $e_1$ and $i$ are shown in Figure \ref{f:jzOICe} and \ref{f:jzOICi}.}
\vspace{0.1cm}
\end{figure*}

Next, we study the surface section in the plane of $J_z-\Omega$ (Figure \ref{f:SSjzO}). These sections clearly show the flip of the orbit when $J_z$ changes sign. The maximum and minimum energy that can be reached in the $J_z-\Omega$ (with $\omega = 0$) plane is $\sim 0$ and $\sim -2.4$. Thus, we plot the surface of section ranging from $H = -2$ to $H = -0.1$ for two values of $\epsilon = 0.001$ and $0.1$. At the quadrupole level, $J_z$ is constant, and there's no resonances in the $J_z-\Omega$ plane. Thus, all the resonances originated from the octupole level of approximation, and the fixed points are at $\Omega = \pi$ and $\Omega = 0$. In addition, similar to the surface section on the $J-\omega$ plane, we see higher order resonances for $\epsilon = 0.1$ at $H=-0.3$ and $H=-0.1$ embedded in the chaotic region, and the chaotic region is confined to $H=-0.5$ and $H=-0.1$. Since $J_z$ changes sign in all energy levels, the orbit may flip for all energy levels, and the flip parameter space is larger for higher $\epsilon$. The corresponding $e_1$ and $i$ on the surface are shown in Figure \ref{f:jzOICe} and Figure \ref{f:jzOICi}. 

To summarize, the surfaces of section show that flips and the excitation of $e_1$ can occur for both regular regions and chaotic regions for a wide range of $H$, and they depend sensitively on the initial condition. In addition, the trajectories are chaotic only when $H\lesssim0$, corresponding to high mutual inclination low eccentricity cases. Furthermore, it is the octupole resonances that cause the flip of the orbit and the excitation of eccentricity very close to unity.

\section{The Maximum Eccentricity and the Flip Condition}
\label{s:EI}
To apply this mechanism to astrophysical systems with different initial conditions, we investigate the parameter regions which exhibit interesting dynamical behaviors. We create a finer grid of $H$ and $\epsilon$ than those presented in Figure \ref{f:SSjo} and \ref{f:SSjzO}, and we monitor the trajectories that start with the selected initial condition in the $J-\omega$ or the $J_z-\Omega$ plane. Of course, some behaviors which do not pass through the selected initial condition will be missed, but this exploration gives a general idea of the behavior of the system as a whole.

We start with the exploration in the $J-\omega$ plane. To systematically estimate the range in $J$ that the trajectories may reach, we start at the maximum energy boundary of $J$ for a given $H$ and $\epsilon$, which corresponds to the minimum eccentricity. Accordingly, for $H < -1$, $\omega$ starts at $\pi$, and for $H>-1$, $\omega$ starts at $\pi/2$. The maximum $e_1$ is recorded after monitoring for $t=500t_K$ (we define $t_K$ in equation (\ref{eqn:tk})), which is much longer than the Lyapunov timescale (see below). 

In Figure \ref{f:CARJ}, we plot $1-e_{1, max}$ as a function of $\epsilon$, where each curve represents a fixed $H$ ($H\in [-2, 2]$), and $\epsilon$ ranges from $0.001$ to $0.1$. In addition, we use the symbol ``x" to mark the $\epsilon$ higher than which the orbit flips. It shows that there are roughly five dynamical regions in $H$: when $H\lesssim -1.5$,  $-0.5\lesssim H\lesssim 0$ and $H\gtrsim 0.5$, the orbit may flip and $e_1$ can be excited very close to unity; when $-1.5\lesssim H\lesssim -0.5$ and $0\lesssim H \lesssim 0.5$, starting with the minimum $e_1$, $e_1$ cannot be excited to unity. Reading from the surface of section in Figure \ref{f:SSjo}, the lack of $e_1$ excitation at $0\lesssim H \lesssim 0.5$ and high $\epsilon$ is due to the quadrupole resonances, which traps the trajectory at low $e_1$.

Particularly, $e_1$ may be excited and the orbit may flip in three scenarios: when the inner orbit is eccentric and coplanar, when the inner orbit is circular and with high inclination, or when the inner orbit is moderately eccentric and with very high inclination $\sim80-90^\circ$ (see Figure \ref{f:JIC}). In addition, the maximum change in $\Delta J$ can be well fit by a power law: 

\begin{eqnarray}
\label{eqn:dJ}
 \Delta J = \left\{ \begin{array}{rl}
	&e^{-2.77H-3.62}\epsilon^{0.051H+1.08} \mbox{ $(H<-1)$} \\
	&e^{2.14H+1.23}\epsilon^{0.75H+2.00}  \mbox{  $(H>-1)$}~,
       \end{array} \right.
\end{eqnarray}

\begin{figure}[h]
\includegraphics[width=3.8in, height=2.3in]{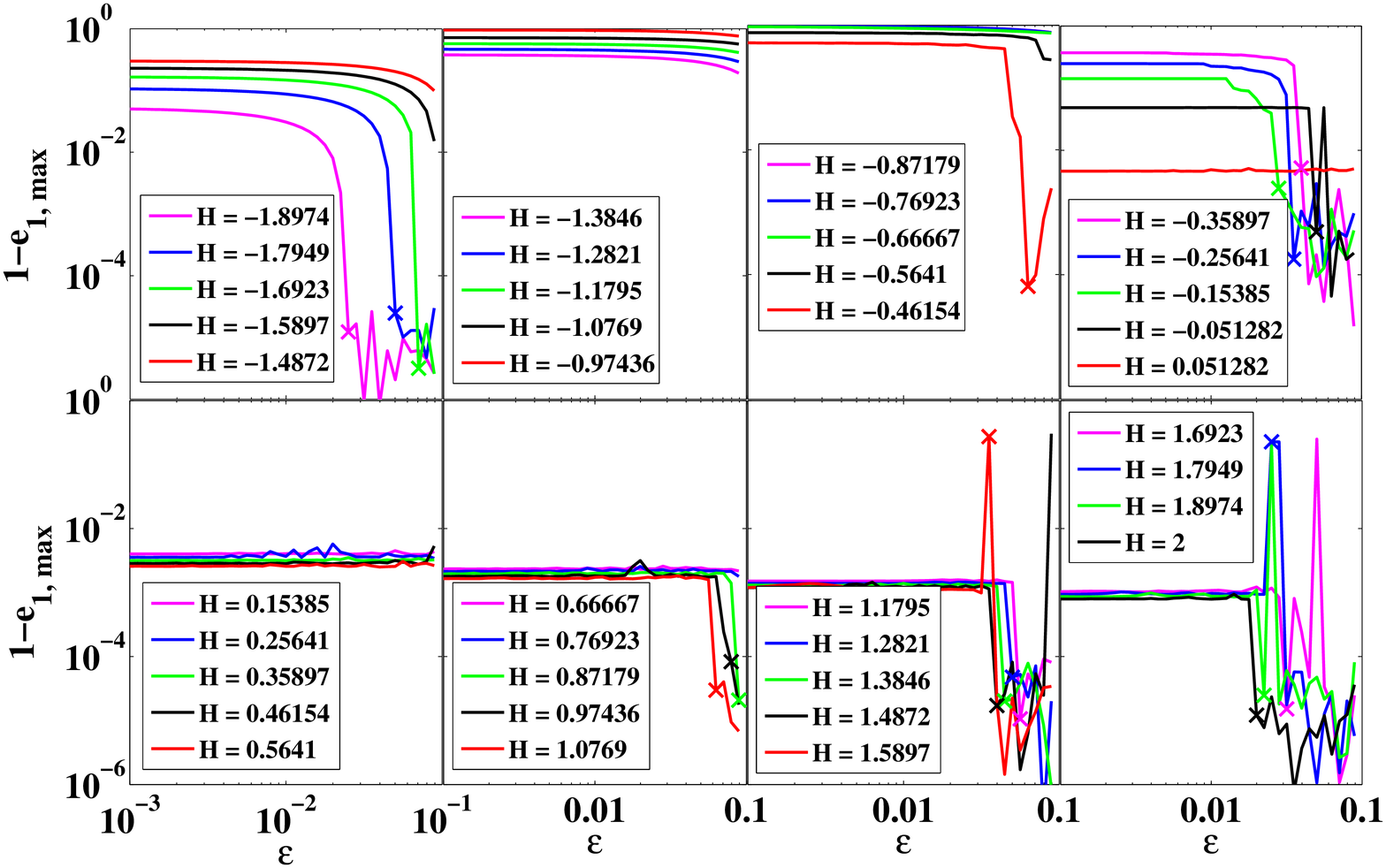}
\caption{\label{f:CARJ} The maximum $e_1$ for different H and $\epsilon$. We set the initial condition at the top of the energy boundary condition of the $J-\omega$ plane, and we record the maximum $e_1$ reached in $t=500t_K$. Each line represents a different H, and the cross marks the $\epsilon$ bigger than which the inner orbit may flip ($i$ cross over $90^\circ$). We find that $e_1$ may be excited and the orbit may flip when$H\lesssim -1.5$,  $-0.5\lesssim H\lesssim 0$ and $H\gtrsim 0.5$. The first case corresponds to the coplanar flip ($i$ flips from $\sim 0^\circ$ to $\sim 180^\circ$ or vise versa), and the latter two correspond to the high inclination flip.}
\vspace{0.1cm}
\end{figure}

\begin{figure}[h]
\includegraphics[width=3.8in, height=2.3in]{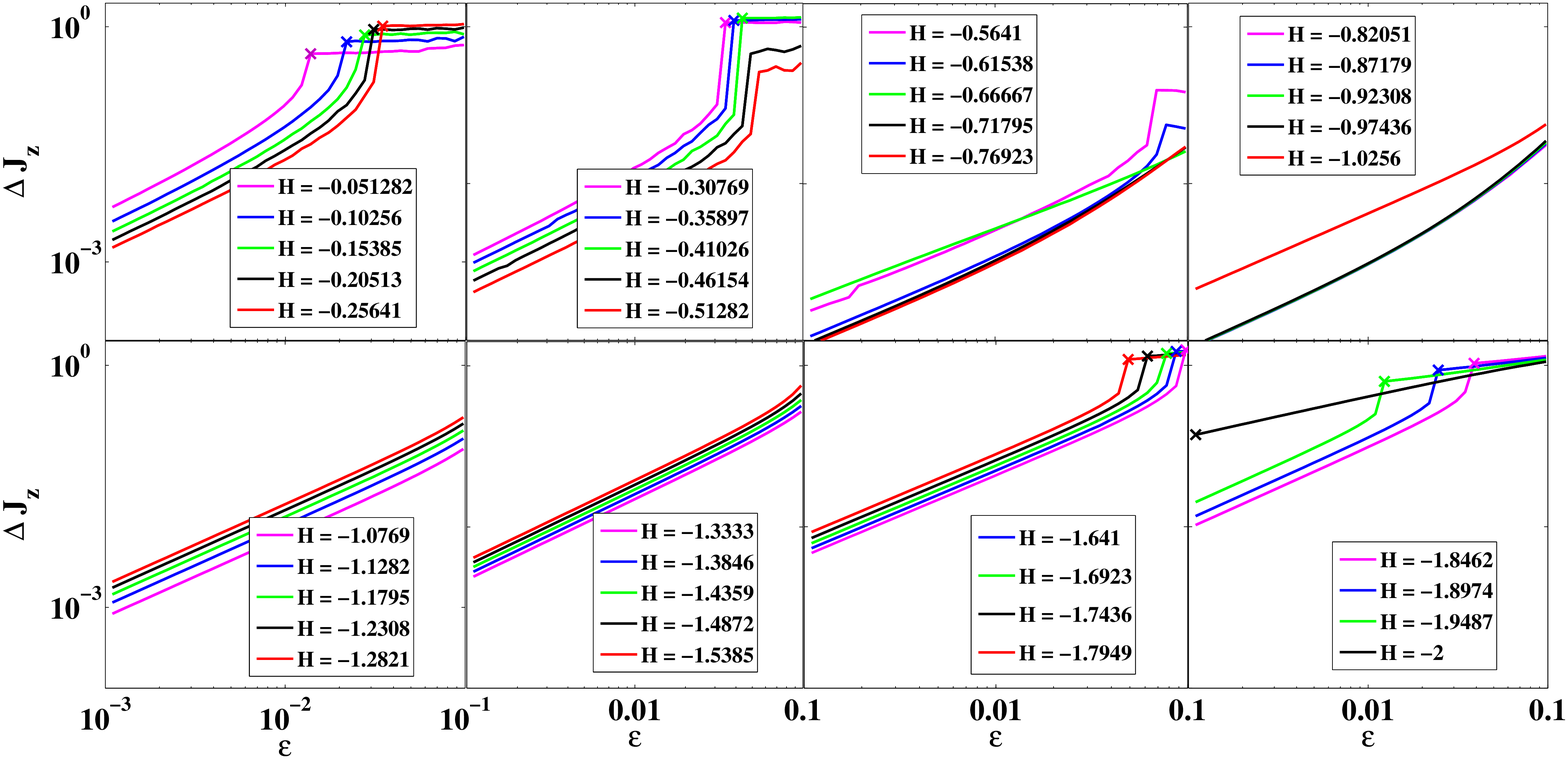}
\caption{\label{f:CARJ_z} The maximum change in $J_z$ for different H and $\epsilon$. We set the initial point at $\Omega = \pi$, where $J_z$ is on the lower energy boundary. We record the maximum change in $J_z$ for $t=500t_K$. The crosses represent the $\epsilon$ bigger than which the inner orbit may flip ($J_z$ changes sign). We find the orbit may flip at $-2<H<-1.5$ and $-0.5<H<0$. The former corresponds to the coplanar flip and the latter corresponds to the high inclination flip.}
\vspace{0.1cm}
\end{figure}

Next, we explore the $J_z-\Omega$ plane. We start the trajectories at the lower energy boundary of $J_z$ at $\Omega = \pi$ for the given $H$ and $\epsilon$, and we record the maximum change in $J_z$ after $t=500t_K$. Figure \ref{f:CARJ_z} shows $\Delta J_z$ as a function of $\epsilon$, where each curve represents a different $H$. $\epsilon$ ranges from $0.001$ to $0.1$, and $H$ ranges from $-2$ to $0$, since the maximum $H$ is zero for $\omega = 0$. Similarly to the $J-\omega$ plane, we use the symbol ``x" to mark the $\epsilon$ higher than which the orbit flips. As expected, it shows that the orbit may flip when $-2<H<-1.5$ and  $-0.5<H<0$, where $-2<H<-1.5$ corresponds to an eccentric and coplanar inner orbit, and $-0.5<H<0$ corresponds to a circular inner orbit with a high inclination. Moreover, $\Delta J_z$ can be fit by a power law of $H$ and $\epsilon$: 
\begin{eqnarray}
\label{eqn:dJz}
 \Delta J_z = \left\{ \begin{array}{rl}
	&e^{-2.21}\epsilon^{1.06}  \mbox{~~~~~~~~~~~~~~~~$(H<-0.5)$} \\
	&e^{10.7H+4.23}\epsilon^{0.48H+1.31}  \mbox{  $(H>-0.5)$}~,
       \end{array} \right.
\end{eqnarray}
Note that equation (\ref{eqn:dJ}) and (\ref{eqn:dJz}) are for the specific initial conditions mentioned above. They show the general dependence of the maximum change in $J$ and $J_z$ on $\epsilon$ and H.

\section{Chaotic Regions}
The surfaces of section show that the system is chaotic when $H \lesssim 0$ (Figure \ref{f:SSjo} and \ref{f:SSjzO}). To better characterize the chaotic regions, we first calculate the percentage of area that is chaotic in each surface in Figure \ref{f:SSjo}. Specifically, we divide each surface into equally spaced grids in $J$ and $\omega$, and count the fraction of grids that has chaotic trajectories. We use the Lyapunov exponent ($\lambda$) to determine whether the trajectories are chaotic, where $\lambda$ indicates how quickly two closely separated trajectories diverge from each other,
\begin{align}
\lambda = \displaystyle \lim_{t \to \infty} \frac{1}{t}\ln\frac{\delta_{traj} (t)}{\delta_{traj} (0)} .
\end{align}

We integrate the tangent of the trajectories for $1000t_K$ to compute $\lambda$, and we find that there are chaotic trajectories only when $\epsilon=0.1$, $H=-0.5$ or $-0.1$. Specifically, $85$ out of $276$ ($\sim 31\%$) grid cells have chaotic trajectories when $\epsilon=0.1$ and $H=-0.5$, and $109$ out of $242$ ($\sim45\%$) grid cells have chaotic trajectories when $\epsilon=0.1$ and $H=-0.1$. It shows even when $H \lesssim 0$, a large range of orbital parameters would still yield regular trajectories.

\begin{figure}
\includegraphics[width=3.5in, height=1.8in]{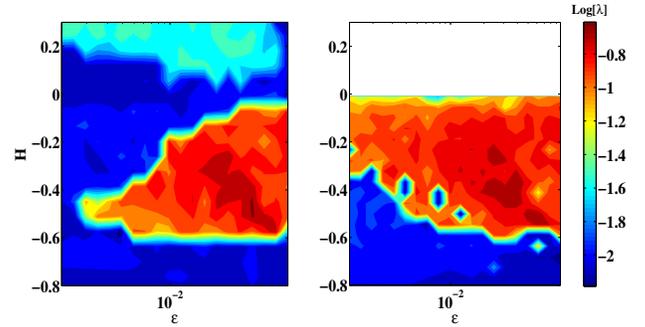}
\caption{\label{f:Lyap} Lyapunov exponents with different $H$ and $\epsilon$. Run time $t=1000$. Left Panel: we consider the following initial condition: $\omega_0=\pi / 2$, $\Omega_0 = 0$, $J_0=1$ or the maximum $J$ at the energy boundary and $-0.8<H<0.3$. Right Panel: we consider the following initial condition: $\omega_0=\Omega_0 = 0$, $J_0=1$ and $-0.6<H<0$. Note that for this choice of initial conditions no physical solution exists for $H>0$. The colormap represents the value of the Lyapunov exponents $\lambda$. The yellow and red colors correspond to big Lyapunov exponents, which are associated with chaotic regions, and cyan and blue colors represent the regular regions.}
\vspace{0.1cm}
\end{figure}

Next, we characterize the chaotic region in the parameter space of $H$ and $\epsilon$. We arbitrarily select the trajectories starting with $\Omega = 0$, $\omega = \pi/2$ and the maximum J for the given $H$ and $\epsilon$, where the associated $e_1$ and $i$ of the initial condition are shown in Figure \ref{f:JIC}. Similarly, we integrate the tangent of the trajectories for $1000t_K$ to compute $\lambda$, and we plot $\lambda$ as a function of $H$ and $\epsilon$ in the left panel of Figure \ref{f:Lyap}. The larger $\lambda$ corresponds to the more chaotic systems. A large region in the parameter space is regular, and the system is chaotic only when $-0.6<H<0$ for larger $\epsilon$. The Lyapunov timescale is $\sim 6t_K$ when $\epsilon \gtrsim 0.01$ and $-0.6<H<0$ (low $e_1$ and $i \gtrsim 40^\circ$). 

To justify that the regions with smaller $\lambda$ are regular, we increase the run time to $4000t_K$, and we find that the Lyapunov exponents for the regular region decrease, while the Lyapunov exponents in the chaotic region remain at $\sim 6t_K$. Moreover, to avoid missing chaotic regions due to the specific choice of the initial condition, we vary the initial condition and make several contour plots of $\lambda$ in the plane of $H$ and $\epsilon$. The right panel of Figure \ref{f:Lyap} shows the case for $\omega$ starts at $0$, where the trajectories are also chaotic when $-0.6<H<0$. 

\section{Conclusion}
The hierarchical three-body system in the test particle limit is common in a large range of astrophysical settings. The dynamical behavior of such systems may lead to retrograde objects, an enhanced rate for tidal disruption, and merger or collision events (e.g. \citet{Holman97, Fabrycky07, Naoz11, Naoz12, Chen11, Bode13}, Naoz \& Silk, in prep, Li et al., in prep). Here, we used a large range of the initial condition to systematically study the dynamics, including the underlying resonances, and the chaotic characteristics of the system.

First, we plotted the surface of section on the $J-\omega$ plane for a large range of energy $H$ and two different $\epsilon$ to identify the underlying resonances (Figure \ref{f:SSjo}). In the quadrupole level, the resonances occur at high $H$ center around fixed points at $\omega = \pi/2$ and $3\pi/2$. On the other hand, the octupole level resonances center at $\omega = 0$, $\pi/2$, $\pi$, or $3\pi/2$ depending on the different energy levels, and we can identify resonances in all these energy levels. The octupole resonances cause the excitation of the $e_1$ in the high eccentricity coplanar case (corresponds to low $H$), shown in \citet{Li13}. The overlap of the quadrupole and octupole resonances causes the chaos for the low eccentricity and high inclination case (corresponds to higher $H$), (e.g. \citet{Naoz11}).

The surfaces of section in the $J_z-\Omega$ plane not only show the octupole resonances but the condition when the orbit flips as $J_z$ changes sign (Figure \ref{f:SSjzO}). At the quadrupole level, $J_z$ is a constant, and there is no resonant zones in the $J_z-\Omega$ plane. However, at the octupole level, the resonant zones exist and lead to the flip of the orbit. As expected, similarly to the $J-\omega$ plane, it also shows that chaotic behavior exist when $H\lesssim0$ for high $\epsilon$. 

Finally, we calculated the Lyapunov exponent for different $H$ and $\epsilon$ to characterize the region where the evolution is chaotic. Consistently with the surface of section, we have found that the orbital evolution is chaotic when $H\lesssim0$ (low $e_1$ high $i$ cases). Specifically, the Lyapunov timescale $\sim 6t_K$.

By monitoring the trajectories, we find that the inner eccentricity may be excited and the orbit may flip for a circular high inclination orbit or for an eccentric and nearly coplanar orbit. This agrees with previous discussions in the literature for the flips with high inclination \citep{Naoz11, Lithwick11, Katz11}, and the coplanar flips \citep{Li13}. In addition, we note that the flips with high inclination are chaotic and the coplanar flips are regular. This analysis can be applied to observed systems. Knowing roughly the orbital elements, one can identify the type of trajectories in the surface of section. Then, one can study the evolution features of the system without doing a large number of simulation for different initial condition. Moreover, our analysis could help predict the enhancement in the rate of tidal disruption events due to eccentricity excitation (Li, et al., in prep).

\acknowledgments
We thank Konstantin Batygin for helpful remarks. This work was supported in part by NSF grant AST-1312034 (for A.L.).

\bibliographystyle{hapj}
\bibliography{msref}

\begin{figure}
\includegraphics[width=6in, height=2.in]{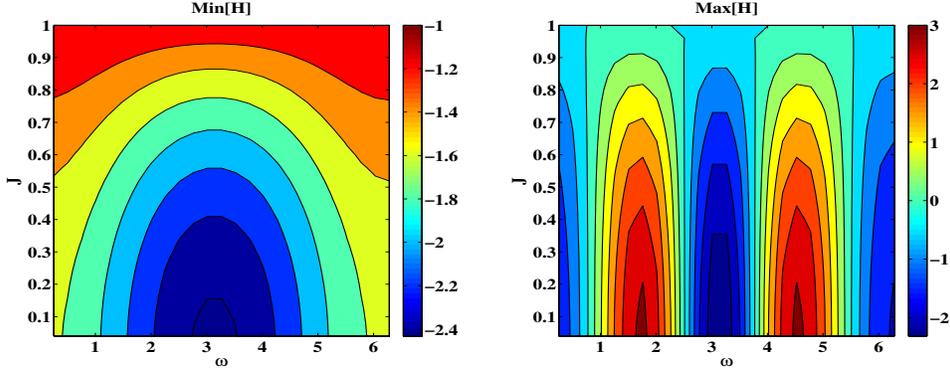}
\caption{\label{f:ei0} The maximum and minimum $H$ it can reach for different $J-\omega$ with $\Omega = 0$. In this plot, we set $\epsilon = 0.1$. The energy range is about $-2.5\sim 3$ in the $J-\omega$ plane. In addition, this explains the shape of the empty region (where there are no solution) in the surface of section plot.}
\label{f:maxminHjo}
\vspace{0.1cm}
\end{figure}

\begin{figure}
\includegraphics[width=6in, height=2.in]{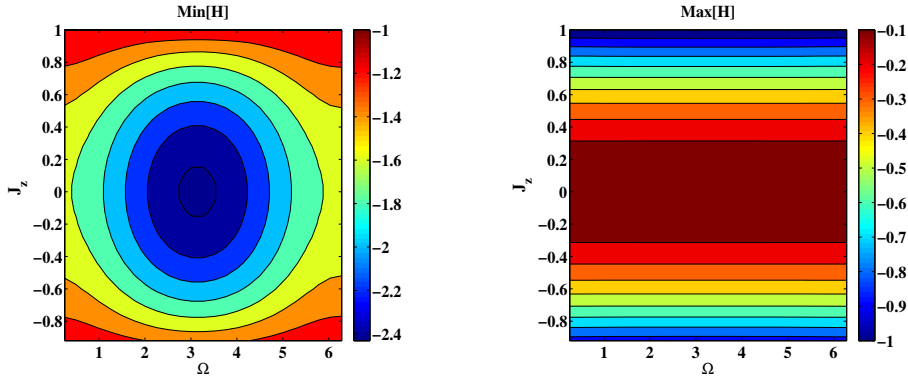}
\caption{\label{f:ei0} The maximum and minimum $H$ it can reach for different $J_z-\Omega$ with $\omega = 0$. In this plot, we set $\epsilon = 0.1$. The energy range is about $-2.5\sim 0$ in the $J_z-\Omega$ plane. In addition, this explains the shape of the empty region (where there are no solution) in the surface of section plot.}
\label{f:maxminHjzO}
\vspace{0.1cm}
\end{figure}

\appendix
First, we explore the range of $H$ it can reach for the surface of section in the $J-\omega$ plane with $\Omega = 0$ and in the $J_z-\Omega$ plane with $\omega=0$.  We contour plot the maximum and minimum of $H$ as a function of $J$ and $\omega$ while setting $\Omega = 0$ in Figure \ref{f:maxminHjo}, which depicts that the range of $H$ is $\sim-2.4$ to $\sim3$. Similarly, we plot the maximum and minimum of $H$ for different $J_z$ and $\Omega$ with $\omega = 0$ in Figure \ref{f:maxminHjzO}. It shows that $H$ ranges from $\sim-2.4$ to $0$. Accordingly, we plot the surface of section for $-2<H<1.2$ in Figure \ref{f:SSjo}, since when $H>1.2$ the section are similar to that when $H=\sim1.2$, and we set $-2<H<0$ for the surface of section in Figure \ref{f:SSjzO}.

Next, we show the associated eccentricity and the inclination for the surface of section (Figure \ref{f:SSjo}, \ref{f:SSjzO}) and the initial condition in Figure \ref{f:CARJ} and \ref{f:CARJ_z}. This helps to connect the resulting dynamical behavior to the parameters in $e_1$ and $i$, that can be obtained more directly for observations. 

\begin{figure}
\includegraphics[width=6in, height=2.7in]{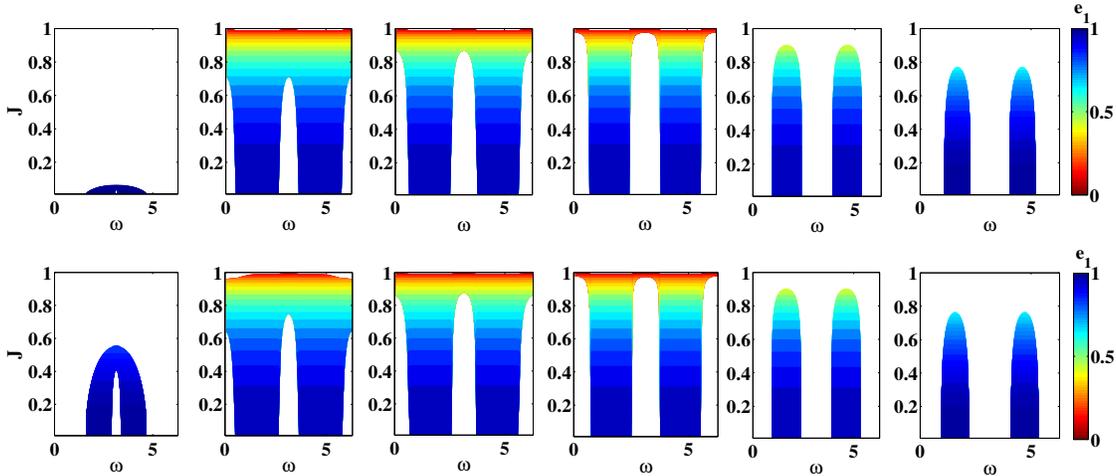}
\caption{\label{f:joICe} The eccentricities in the $J-\omega$ plane ($\Omega = 0$). Note that these are not the initial conditions, but directly the values of $e_1$ in the $J-\omega$ surface at fixed $\Omega=0$ for the given $H$ and $\epsilon$. Similar to Figure \ref{f:SSjo}, in the first row, $\epsilon = 0.001$ and in the second row, $\epsilon = 0.1$. The octupole terms are more dominant when $\epsilon$ is bigger. }
\vspace{0.1cm}
\end{figure}

\begin{figure}
\includegraphics[width=6in, height=2.7in]{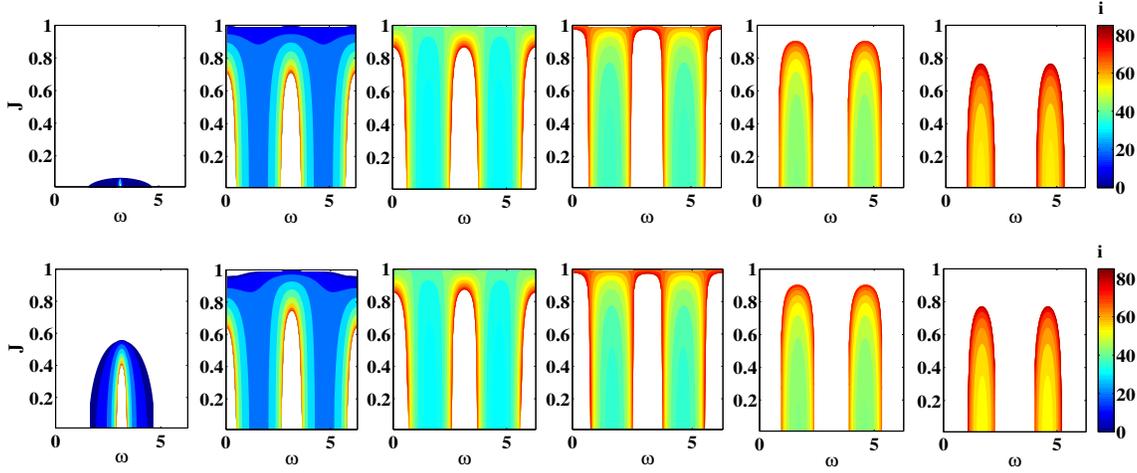}
\caption{\label{f:joICi} The inclinations in the $J-\omega$ plane ($\Omega = 0$). Note that these are not the initial conditions, but directly the values of $i$ in the $J-\omega$ surface at fixed $\Omega=0$ for the given $H$ and $\epsilon$. Similar to Figure \ref{f:SSjo}, in the first row, $\epsilon = 0.001$ and in the second row, $\epsilon = 0.1$. The octupole terms are more dominant when $\epsilon$ is bigger. }
\vspace{0.1cm}
\end{figure}

In Figure \ref{f:joICe} and \ref{f:joICi}, we plot the initial condition in the $J-\omega$ plane corresponding to the surfaces of section in Figure \ref{f:SSjo}. $e_1$ can be calculated from the $J$ value directly as $e_1 = \sqrt{1-J^2}$, so higher $J$ associates with lower $e_1$. On the other hand, $i$ is lower for larger $J$ when $H = -2, -1, -0.5$, and $i$ is higher for larger $J$ when $H = -0.1, 0.5, 1.2$.

\begin{figure}
\includegraphics[width=6in, height=2.7in]{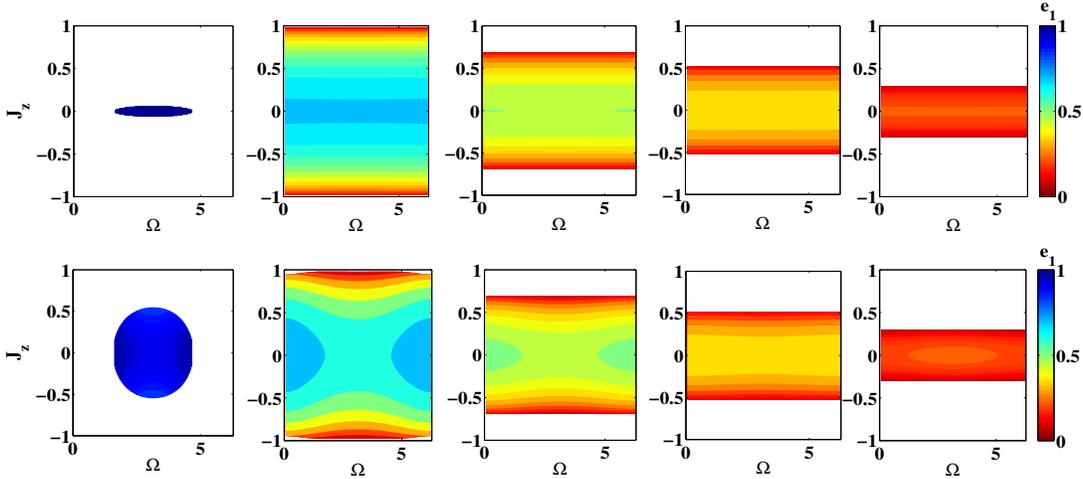}
\caption{\label{f:jzOICe} The eccentricities on the $J_z-\Omega$ plane ($\omega = 0$). Note that these are not the initial conditions, but directly the values of $e_1$ in the $J_z-\Omega$ surface at fixed $\omega=0$ for the given $H$ and $\epsilon$. Similar to Figure \ref{f:SSjzO}, in the first row, $\epsilon = 0.001$ and in the second row, $\epsilon = 0.1$. The octupole terms are more dominant when $\epsilon$ is bigger. }
\vspace{0.1cm}
\end{figure}

\begin{figure}
\includegraphics[width=6in, height=2.7in]{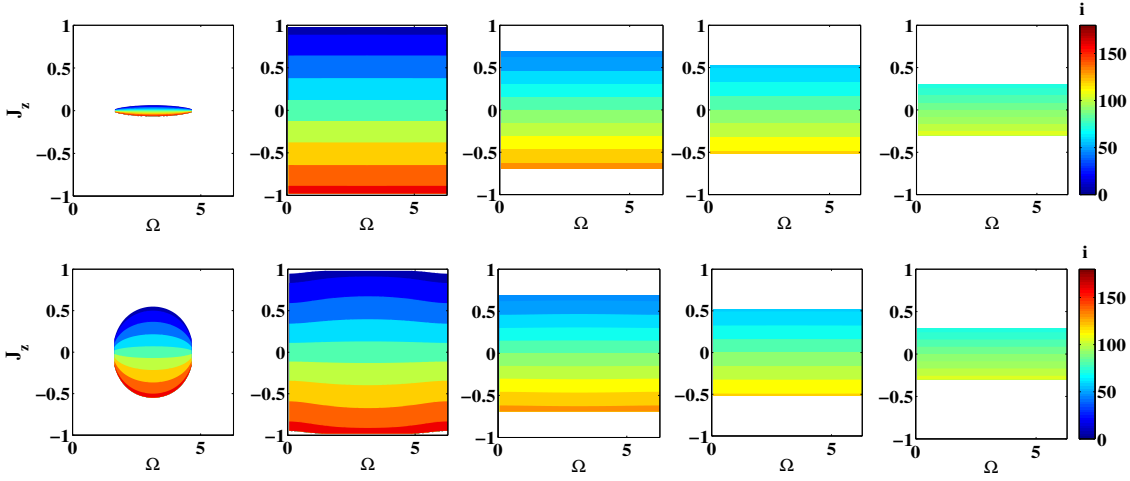}
\caption{\label{f:jzOICi} The inclinations in the $J_z-\Omega$ plane ($\omega = 0$). Note that these are not the initial conditions, but directly the values of $i$ in the $J_z-\Omega$ surface at fixed $\omega=0$ for the given $H$ and $\epsilon$. Similar to Figure \ref{f:SSjzO}, in the first row, $\epsilon = 0.001$ and in the second row, $\epsilon = 0.1$. The octupole terms are more dominant when $\epsilon$ is bigger. }
\vspace{0.1cm}
\end{figure}

Next, in Figure \ref{f:jzOICe} and \ref{f:jzOICi}, we plot $e_1$ and $i$ in the $J_z-\Omega$ plane, corresponding to the surface section in the $J_z-\Omega$ plane with $\omega=0$ in Figure \ref{f:SSjzO}. When $i>90^\circ$, $J_z > 0$, and when $i<90^\circ$, $J_z < 0$. We find that $e_1$ is higher for lower $H$, and $i$ is closer to $90^\circ$ for higher $H$.

\begin{figure}
\includegraphics[width=6in, height=2.5in]{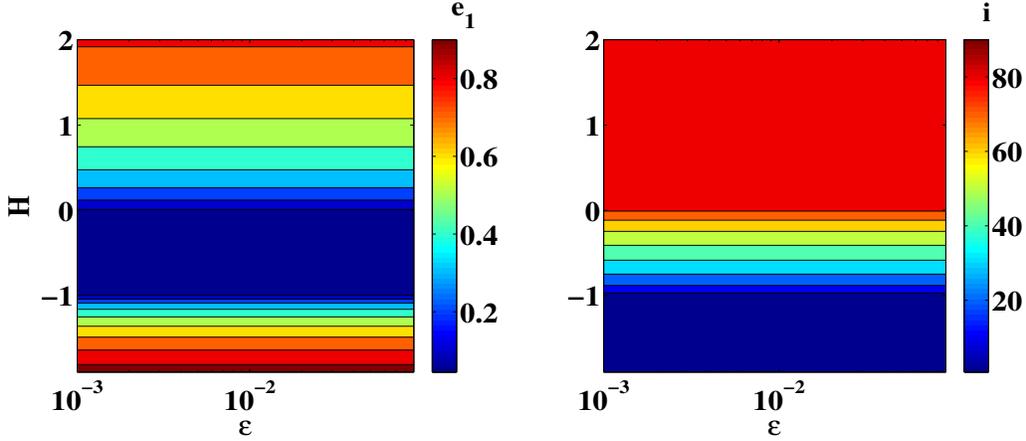}
\caption{\label{f:JIC} $e_1$ and $i$ with $\omega_0 = \pi$ ($H<-1$), $\omega_0=\pi/2$ ($H>-1$), $\Omega=0$ and $J$ at the upper energy boundary. This is associated with the initial condition for Figure \ref{f:CARJ}, i.e. the initial $e_1$ and $i_1$ for each run at fixed $\epsilon$ and $H$.}
\vspace{0.1cm}
\end{figure}

Furthermore, we plot the initial condition for the trajectories we selected to investigate the maximum $e_1$ in Figure \ref{f:JIC}. It shows that for the maximum $e_1$ plot (Figure \ref{f:CARJ}), when $H\lesssim-1.2$, we monitor the trajectories that start with high eccentricity and low inclination. In this case, when $H\lesssim-1.7$, the orbit may flip at high $\epsilon$ and the maximum $e_1$ may reach $\sim 1-10^{-6}$ for high $\epsilon$. When $-1.2\lesssim H \lesssim 0$, we monitor trajectories that start with low eccentricity and high inclination. In this case, not much variations are seen unless $H\lesssim0$. When $H>0$, we monitor trajectories starting with high inclination $i\sim80 - 90^\circ$.

\begin{figure}
\includegraphics[width=6in, height=2.7in]{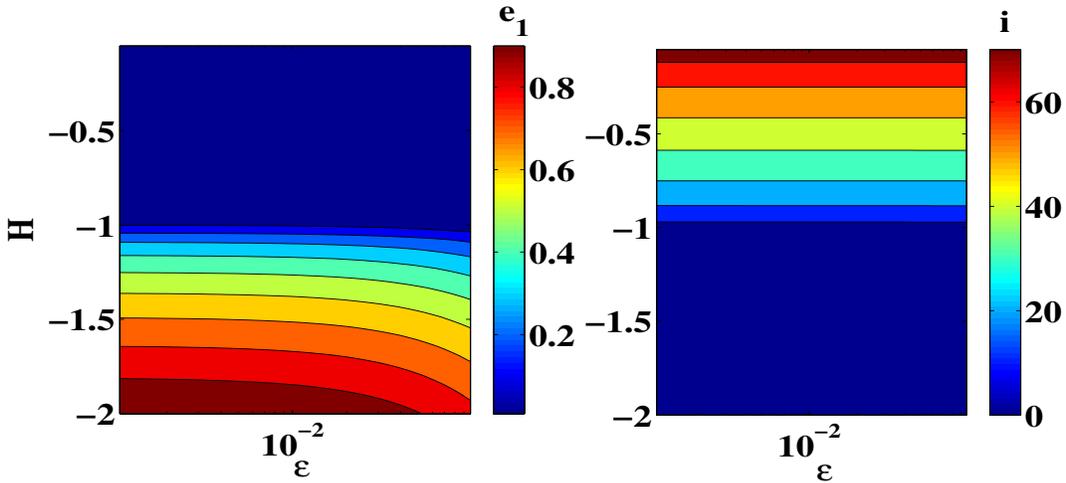}
\caption{\label{f:J_zIC} $e_1$ and $i_1$ with $\Omega = \pi$, $\omega = 0$ and $J_z$ are at lower energy boundary for different $H$ and $\epsilon$. This is associated with the initial condition in Figure \ref{f:CARJ_z}, i.e. the initial $e_1$ and $i_1$ for each run at fixed $\epsilon$ and $H$.}
\vspace{0.1cm}
\end{figure}

In the end, we plot the initial condition for the trajectories that are monitored for $J_z$ or the flip of the orbit in Figure \ref{f:J_zIC}. When $H\lesssim-1$, we start the trajectories with high $e_1$ and low $i$; when $H\gtrsim -1$, we start the trajectories with low $e_1$ and high $i$. The orbit may flip with $-0.4\lesssim H\lesssim0$ at high $\epsilon$ for trajectories starting with low $e_1$ and high $i$, and the orbit may flip with $H\lesssim -1.5$ when the trajectories start with high $e_1$ and low $i$.

\end{document}